\PassOptionsToPackage{square, comma, sort&compress}{natbib}
\documentclass[preprint,12pt]{elsarticle}

\usepackage{amsmath}
\usepackage[colorlinks=true,linkcolor=blue,urlcolor=blue,citecolor=blue]{hyperref}
\usepackage{graphicx}
\usepackage[squaren, Gray, cdot]{SIunits}
\usepackage{amssymb}
\usepackage{mathtools}

\usepackage{tikz}
\journal{Physics Letters A}

\begin{document}

\begin{frontmatter}



\title{Observation of Magnetically-Induced atomic transitions of the Cs 6S$_{1/2} \rightarrow 7$P$_{3/2}$
line at 456 nm
}


\author[IPR]{Armen Sargsyan}
\author[FEMTO]{Arevik Amiryan}
\author[FEMTO]{Emmanuel Klinger\corref{mycorrespondingauthor}}
\cortext[mycorrespondingauthor]{Corresponding author}
\ead{emmanuel.klinger@femto-st.fr}
\author[IPR]{David Sarkisyan}

\address[IPR]{Institute for Physical Research – National Academy of Sciences of Armenia, Ashtarak-2 0203, Armenia}

\address[FEMTO]{Universit\'e Marie et Louis Pasteur, Supmicrotech, CNRS, Institut FEMTO-ST, F-25000 Besan\c con, France}


\begin{abstract}


It has recently been demonstrated that magnetically induced (MI) transitions, a class of transitions forbidden at zero magnetic field, of the  Cs 6$^2$S$_{1/2} \rightarrow 6^2$P$_{3/2}$ (D$_2$) line, exhibit promising features for high-resolution physics applications in the near-infrared range. In this work, we study a group of seven MI transitions ($F_g = 3 \rightarrow F_e = 5$) of the  Cs $6^2$S$_{1/2} \rightarrow 7^2$P$_{3/2}$ line at $\lambda = 456$\;nm.  The experimental measurements are in very good agreement with theoretical predictions based on the diagonalization of the Zeeman Hamiltonian. In magnetic fields ranging from $0.2-3$\;kG, these transitions reach a maximum intensity above that of conventional transitions. Another noteworthy property is their large frequency shift, reaching approximately $17~\mathrm{GHz}$ with respect to the unperturbed hyperfine transitions in magnetic fields of about $3~\mathrm{kG}$. These interesting properties may prove useful for the realization of optical frequency references or magnetometers with sub-micron spatial resolution in the blue region of the spectrum.

\end{abstract}




\end{frontmatter}

\section{Introduction}
 
Magneto-optical processes are widely used in many areas as they encompass a large range of applications, such as optical magnetometry \cite{FabricantNJP2023}, narrow-band atomic filtering \cite{Uhland2023}, tunable laser frequency locking \cite{Musezahl2024}, etc. For this reason, the interaction of light with an atomic vapor perturbed by a magnetic field has been widely studied and is still a topic of interest \cite{tremblayPRA1990,scottoPRA2015,staerkind2023high,scottoEPJD2026}. Recently, attention has been focused on atomic transitions between ground and excited hyperfine levels satisfying an apparent $F_e - F_g = \Delta F = \pm 2$ selection rule ~\cite{momier2021JQSR, tonoyan2023formation, sargsyanPhysLettA2025, hanPhysRevA2025,tianPhysLettA2025}. 
The intensity of these transitions is zero when no external magnetic field is applied but significantly increases with the magnetic field amplitude. These transitions are referred to as magnetically induced (MI) transitions. 

\begin{figure}[htb]
    \centering
    \includegraphics[width=0.9\textwidth]{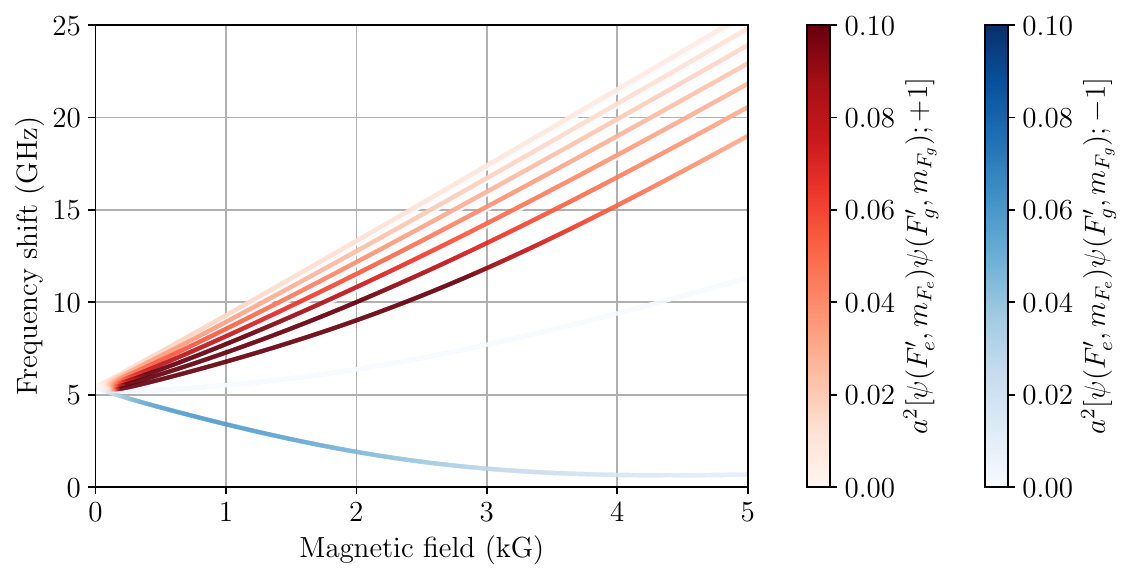}
    \caption{
    Evolution of $F=3\rightarrow 5'$ MI transitions of the Cs D$_2$ line in a magnetic field. The plot shows the evolution of the transition frequency as a function of the magnetic field amplitude. Here, the zero frequency corresponds to the weighted center of the D$_2$ line, i.e. $\nu_{\text{D}_2} \approx 351.725 719$\;THz. The line colors reflect the evolution of the transition intensity, through the transfer coefficient $a^2[\psi(F_e',m_{F_e})\psi(F_g',m_{F_g});q]$ in the case of $\sigma^+$ ($q=+1$, red) and $\sigma^-$ ($q=-1$, blue) optical polarization of the resonant light. The theoretical model used to calculate the behavior of atomic transitions in an external static magnetic is based on constructing the magnetic Hamiltonian \cite{tremblayPRA1990}, including the entire hyperfine manifold in the $\lvert F, m_F \rangle$ basis, and then perform numerical diagonalization, see Sec.\;\ref{sec:theory}.}
    \label{fig:Fig1}
\end{figure}

Important peculiarities have been demonstrated for the group of seven $F=3 \rightarrow 5'$ (where the prime indicates the excited state)
MI transitions of the $6^2\text{S}_{1/2} \rightarrow 6^2\text{P}_{3/2}$ (D$_2$) transition at $\lambda = 852~\mathrm{nm}$ of Cs atoms~\cite{tonoyanEPL2018}. The intensities of MI transitions with $\Delta F = +2$ are maximal (and the number of MI transitions is also maximal) for $\sigma^{+}$ circularly polarized laser radiation, whereas the intensities of MI transitions with $\Delta F = -2$ are minimal (and the number of MI transitions is also minimal) for $\sigma^{-}$ circularly polarized radiation, see Fig.\;\ref{fig:Fig1}.
This demonstrates a remarkable difference in the behavior of the $F = 3 \rightarrow 5'$ transitions depending on the use of right- or left-handed circularly-polarized radiation.
An attractive feature of these MI transitions is their large frequency shift with respect to the unperturbed hyperfine transitions in a magnetic field, which reaches approximately $17~\mathrm{GHz}$ at a magnetic field of $B \approx 3~\mathrm{kG}$. 
Moreover, it is important to note that these transitions are formed on the high-frequency wing of the spectrum and do not overlap with other transitions, which is of practical interest.
For example, magnetically induced transitions of the Cs D$_2$ line have been successfully used for the realization of electromagnetically induced transparency \cite{sargsyan2022coherent}.

In this article, we report on a combined experimental and theoretical study of $F=3 \rightarrow 5'$ MI transitions of the $6^2\text{S}_{1/2} \rightarrow 7^2\text{P}_{3/2}$ line at 456\;nm. This is, to the best of our knowledge, the first time these transitions are studied.
Indeed, performing such a study is quite a challenge for the following reasons. $(i)$ The frequency spacing between the upper levels of $7^2\text{P}_{3/2}$ is approximately 3 times smaller than that of the $6^2\text{P}_{3/2}$ of the D$_2$ line, $\lambda = 852$\;nm \cite{Williams2018}. $(ii)$ The Doppler width for the $6^2\text{S}_{1/2} \rightarrow 7^2\text{P}_{3/2}$ transition ($\Gamma_D/2\pi = 0.87$\;GHz at $180\,^\circ$C) is approximately 2.5 times larger than that for the D$_2$ line, which greatly complicates the study of individual atomic transitions, especially in the presence of a magnetic field. $(iii)$ The oscillator strength
is almost two orders of magnitude smaller here than for the transition to the $6^2\text{P}_{3/2}$  state \cite{Safronova2016,Toh2019}. For these reasons, spectroscopic studies of the $6^2\text{S}_{1/2} \rightarrow 7^2\text{P}_{3/2}$ transition are scarce in the literature. Previous studies have focused on the observation of these transitions in nano- \cite{sargsyanJPhysB2025,sargsyan2025OptLett} and millimeter-sized \cite{Klinger2024Sub-doppler} vapor cells and their application for laser frequency stabilization \cite{Miao2022,Klinger2025ApplPhysLett}. To study a large number closely-spaced atomic transitions split by a magnetic fields, a sub-Doppler technique is thus required.

As was previously shown, using nanocell combined with derivative spectroscopy allows one to study individual atomic transitions even in the presence of a large number of Zeeman components \cite{sargsyanPhysLettA2025}. To spectrally separate MI transitions from other atomic transitions that form in the spectrum in a magnetic field, we used a process referred to as selective reflection (SR). The SR of laser radiation is formed at the boundary between an alkali metal vapor and dielectric windows and is known to provide nearly Doppler-free spectral resolution, see e.g. Refs.~\cite{weis1992AmPhysSoc,vartanyan1959PhysRevA,dutierJOSAB2003}, usually employed to study atom-surface interactions \cite{failache1999PhysRevLett,Laliotis2021atom-surface} or spectral features in high density vapor cells \cite{sautenkov2024_SRpumpProbe,sautenkov2026_SRnonlinear}.

The manuscript is organized as follows. In Sec.\;\ref{sec:theory}, we briefly review how to compute the evolution of transition frequencies and intensities in a magnetic field and show the prediction for the $6^2\text{S}_{1/2} \rightarrow 7^2\text{P}_{3/2}$ transition. Sec.\;\ref{sec:setup} is dedicated to detailing the experimental setup. Finally, in Sec.\;\ref{sec:results} we compare experimental and theoretical results.

\section{Theoretical considerations}\label{sec:theory}

To calculate the evolution of alkali transitions in an external magnetic field, we follow the procedure detailed in Ref.\;\cite{tremblayPRA1990}. Briefly, this consists in diagonalizing the  Hamiltonian matrix which accounts for the hyperfine atomic structure 
and the Zeeman Hamiltonian
\begin{equation}\label{eq:ZeemanMatrix}
    H = H_{\text{hfs}} +\frac{\mu_B}{\hbar}(g_S\,S_z +g_L\,L_z + g_I\,I_z) B_z\,,
\end{equation}
where $\mu_B$ is the Bohr magneton \cite{olsen2011optical}, $g_{S,L,I}$ are the Landé factors \cite{staerkind2023precision} and $S_z$, $L_z$, $I_z$ the projection of quantum numbers on the $z$-axis (quantization axis). Expressions for diagonal and non-diagonal matrix elements in the base $|F,m_F\rangle$ are given in Ref.\;\cite{tremblayPRA1990}. Diagonalizing this matrix provides access, on the one hand, to the eigenvalues directly linked to the resonance frequencies. On the other hand, the transition intensities are proportional to the squared dipole moment, given, in the coupled basis $|F,m_F\rangle$, by 
\begin{equation}
    \left\lvert \left\langle e \lVert d\lVert g\right\rangle\right\lvert^2 = \frac{3\epsilon_0\hbar \lambda_{eg}^3}{8 \pi^2} \cdot \Gamma_\text{N} \, a^2[\psi(F_e',m_{F_e});\psi(F_g',m_{F_g});q]\,,
\end{equation}
where $\Gamma_\text{N}/2\pi \approx 1.16\,$MHz \cite{Toh2019} and $\lambda_{eg}$ are respectively the natural linewidth and the wavelength of the $6^2\text{S}_{1/2} \rightarrow 7^2\text{P}_{3/2}$ transition. The transfer coefficients $a[\psi(F_e',m_{F_e});\psi(F_g',m_{F_g});q]$ modified by the field  are calculated using the eigenvectors. Here, $q=0,\pm1$ is associated with the polarization of the incident laser field. This computation is repeated for every value of the magnetic field to construct the evolution of the transitions. Note that, in the case of the alkali D lines, these can be computed with \textit{ElecSus} \cite{zentile2015elecsus}.

 \begin{figure}[htb]
    \centering
    \includegraphics[width=0.95\textwidth]{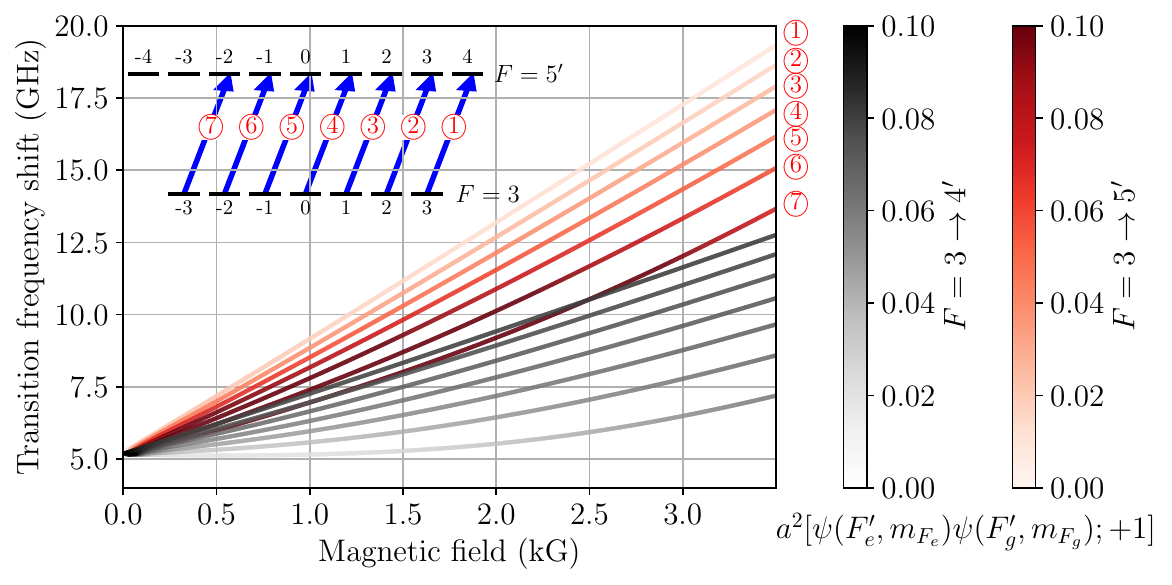}
    \caption{Evolution of $F=3\rightarrow 5'$ MI transitions (red) and $F=3\rightarrow 4'$ (black) of the Cs  $6^2\text{S}_{1/2}\rightarrow7^2\text{P}_{3/2}$ line in a magnetic field. The plot shows the evolution of the transition frequency as a function of the magnetic field intensity. Here, the zero frequency corresponds to the weighted center of the D$_2$ line, i.e. $\nu_{6\text{S}_{1/2}\rightarrow7\text{P}_{3/2}} \approx 657.936362$\;THz \cite{Williams2018}. The line colors reflect the evolution of the transition intensity, through the transfer coefficient $a^2[\psi(F_e',m_{F_e})\psi(F_g',m_{F_g});q]$, in the case of $\sigma^+$ optically-polarized resonant light ($q=+1$). The inset shows a diagram of the transitions with the corresponding labeling.}
    \label{fig:Fig2}
\end{figure}

In Fig.\;\ref{fig:Fig2}, we plot both the dependence of the frequency and intensity of the $F=3 \rightarrow 5'$ group of transitions under $\sigma^+$ laser excitation as a function of the external magnetic field $B$. This group contains seven magnetic field-induced atomic transitions labeled \textcircled{1} to \textcircled{7}, see the inset in Fig.\;\ref{fig:Fig2}. The plot also includes the group of $F=3\rightarrow4'$ transitions as these two groups partially overlap at low and intermediate magnetic fields. To qualify these regimes, it is often convenient to introduce the characteristic quantity
\begin{equation*}
B_0 = \frac{A_{\mathrm{HFS}}}{\mu_B} \approx 1.6\;\text{kG},
\end{equation*}
where $A_{\mathrm{HFS}}$ is the hyperfine structure coupling coefficient for the Cs ground level $6^2\text{S}_{1/2}$. This quantity describe how strong the interaction of the atoms with the magnetic field is: when $B\ll B_0$, the field is described as low. Conversely, when $B\gg B_0$ the field is described as high and, in between, the field is qualified as intermediate.

In the range of $0.1$--$3.4~\mathrm{kG}$, MI transitions are the strongest atomic transitions among all transitions originating from the $F_g = 3$ ground state. Note that the highest intensity corresponds to the MI transition for which the magnetic sublevel of the ground state $m_F$ is minimal ($m_F=-3\rightarrow-2'$); in this case \textcircled{7}. Conversely, the smallest intensity corresponds to transition \textcircled{1}, for which the value of $m_F$ is the largest ($m_F=3\rightarrow 4'$). For $B \gg B_0$, our calculations show that the intensity of these MI transitions tends to zero.
In addition, for magnetic fields $B > 3~\mathrm{kG}$
 these transitions are formed on the high-frequency wing of the spectrum and do not overlap with others.
This behavior is due to the fact that the frequency slope of the $F= 3 \rightarrow 5'$ MI transitions for $B > 2~\mathrm{kG}$ is approximately $4~\mathrm{MHz/G}$, while for $F= 3 \rightarrow 4'$ transitions, the frequency slope is approximately $2~\mathrm{MHz/G}$.

Note that there also exist five MI transitions of the group $F= 2 \rightarrow 4'$ (not shown in Fig.\;\ref{fig:Fig2}), as they are only excited with $\sigma^-$ circularly-polarized laser radiation. However, these transitions have intensities much weaker than the intensities of the $F= 3 \rightarrow 5'$ group as in the case of the D$_2$ line \cite{sargsyan2022coherent}.
Additionally, there exists a set of MI transitions of the type
\begin{equation*}
   |F_g = F,\, m_F = 0\rangle \rightarrow |F_e = F,\, m_F' = 0\rangle 
\end{equation*}
which are forbidden in zero magnetic field, but can be excited with linearly $\pi$-polarized laser excitation when $B > 0$ \cite{sargsyanJPB2020}. It is interesting to note that for $B \gg B_0$, these MI transitions tend toward an asymptotic value.
In summary, our simulations show a total of 17 MI transitions for the $6^2\text{S}_{1/2} \rightarrow 7^2\text{P}_{3/2}$ transition, which expand the possibilities for using magneto-optical processes in the blue spectral region. For the experimental study, we will focus only on the $F= 3 \rightarrow 5'$ MI transitions.

\section{Experimental Setup}\label{sec:setup}

\begin{figure}[htb]
    \centering
    \includegraphics[width=0.55\textwidth]{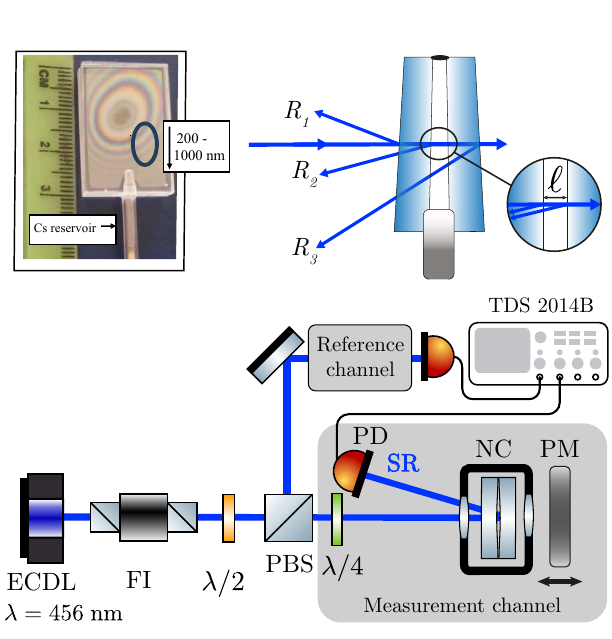}
    \caption{Sketch of the experimental setup. ECDL -- extended cavity diode laser operating at 456 nm, FI -- Faraday isolator, PBS -- polarizing beam splitter, PM -- permanent magnet, NC -- nanometric-thin cell filled with Cs in the oven, PD -- photodiodes, TDS 2014B -- Digital oscilloscope, $\lambda/2$, $\lambda/4$-- half and quarter wave plates. The PM is oriented such that the applied magnetic field \textbf{B} is collinear with the propagation axis ($z$-axis) of the laser radiation (\textbf{B}$\parallel$\textbf{k}) in order to excite $\pi$ transitions. The reference channel comprises  d. Upper left inset, is the photograph of the NC; the oval marks 200–1000\;nm region of the Cs vapor column. Upper right inset, geometry of the three reflected beams. The SR beam propagates in the direction of $R_2$ and was used to realize sub-Doppler spectroscopy.}
    \label{fig:Fig3}
\end{figure}

In order to resolve Doppler-broadened, closely-spaced, atomic transitions we use selective reflection from a nanocell (NC) combined with derivative spectroscopy. The main advantages of SR from a NC are the following: $(i)$ simplicity of the realization (single beam, as opposed to counter-propagating beams required for e.g. saturated absorption); $(ii)$ absence of crossover resonances, which is particularly important when studying atomic transitions in an external field as CO resonances splits into many additional components; $(iii)$ the amplitudes of the formed atomic transitions are correctly related according to their theoretical intensities. 

The sketch of the experimental setup is presented in Fig.\;\ref{fig:Fig3}. We use an external-cavity diode laser (ECDL) with a spectral linewidth of 400\;kHz tunable around $\lambda = 456$ nm and resonant with the $6^2\text{S}_{1/2} \rightarrow 7^2\text{P}_{3/2}$ transition of Cs. To avoid feedback in the laser cavity, a Faraday insulator (FI) was used. The incident laser beam is then directed at normal incidence onto the nanocell windows. A photography of the NC (front view) is shown on the left inset of Fig.~\ref{fig:Fig3} where we have circled the 200-1000\;nm range of thicknesses. The NC consists of $\sim$2 mm-thick windows made of technical sapphire (Al$_2$O$_3$) and a vertical side arm (a sapphire tube reservoir  filled with Cs metal). The laser beam configuration for SR spectroscopy with the NC  is depicted in right inset of Fig.~\ref{fig:Fig3}. Note that the wedge of the nanocell windows are exaggerated on this diagram.  To record the signal of interest ($R_2$), arising from reflections of the window-vapor and vapor-window interfaces inside the cell, the reflected beam is detected with a photodiode whose signal is fed to a four-channel oscilloscope (Tektronix TDS2014B). 

A part of the laser radiation was directed to an auxiliary unit to provide a frequency reference based on saturated absorption \cite{Li2019OptLett}. To study MI transitions, a strong permanent neodymium magnet (PM) was mounted on a micrometric-step translation stage in the proximity of the cell’s rear window. The $B$-field strength was varied by simple longitudinal displacement of the magnet. In this configuration, the magnetic field is directed along the laser radiation propagation axis ($z$, the quantization axis) such that $\mathbf{B} \parallel \mathbf{k}$. The applied $B$-field was calibrated with a Teslameter HT201 magnetometer.

 To overcome smaller signals due to the weaker dipole moments (with respect to the D$_2$ line), we heat the NC to a temperature of about 180\,$^\circ$C while the windows were maintained at a temperature that was about 20\,$^\circ$C higher to avoid condensation of the atomic vapor on the windows. The laser is shined on the region $\ell \approx 800$ nm. While this is not optimal for performing high resolution spectroscopy with nanocells \cite{dutier2003collapse}, it allows recording resonances with a better signal-to-noise ratio while minimizing the contribution of van der Waals interactions that would be present at lower thicknesses \cite{Sargsyan2026features}. The laser power was set to about 30\;mW  (beam diameter of $\approx$ 1 mm). Despite this power being high, it was shown to not be saturating in the case of the Cs D$_2$ line recorded with a NC \cite{andreeva2007PhysRevA}.


\section{Selective reflection spectra in an external magnetic field}\label{sec:results}

\begin{figure}[ht]
    \centering
    \includegraphics[width=0.95\textwidth]{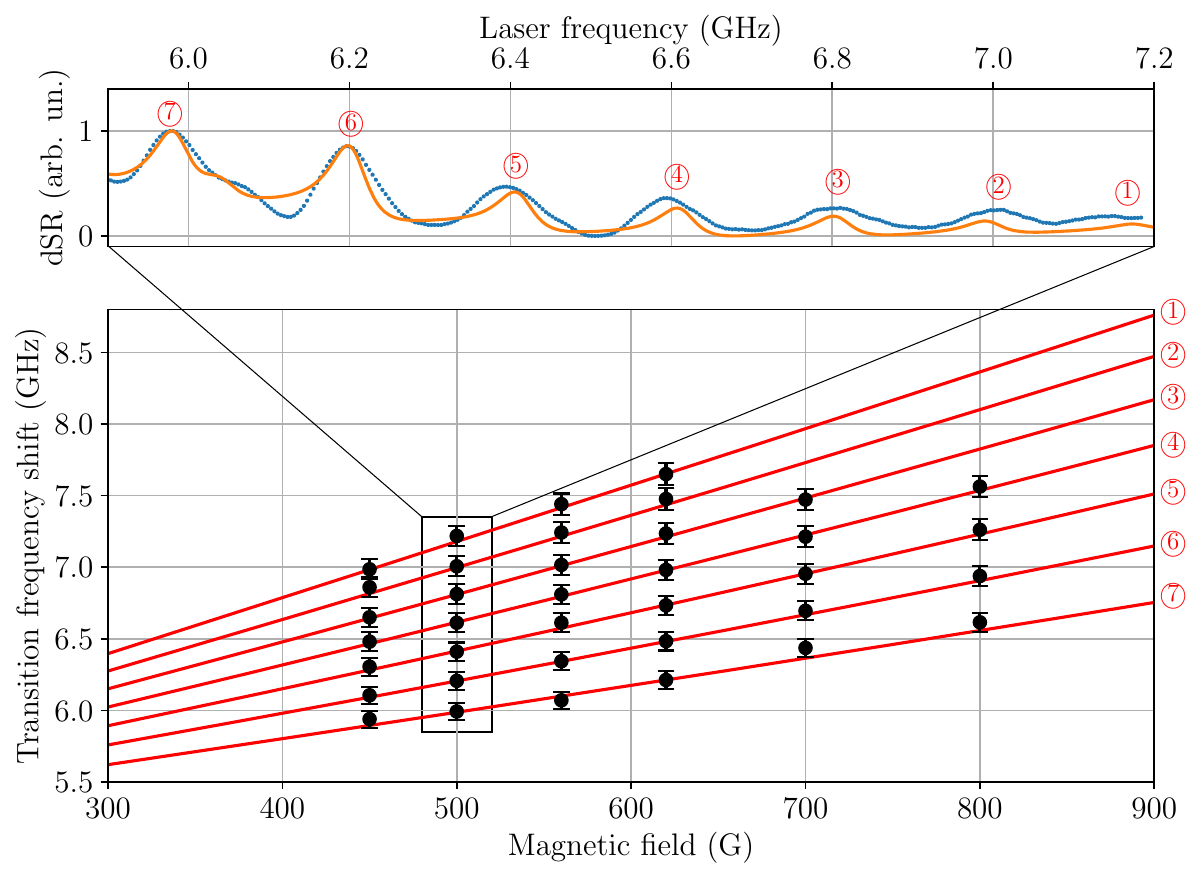}
    \caption{Magnetic dependence of the $6^2\text{S}_{1/2}\rightarrow7^2\text{P}_{3/2}$ $F=3\rightarrow5'$ MI transitions labeled \textcircled{1} to \textcircled{7}. The dots correspond to the experimental values extracted from the spectrum and the error bars correspond to the each observed transition linewidth. The red solid line correspond to the expected theoretical evolution following the model described in Sec.\;\ref{sec:theory}. The top inset shows an example of experimental derivative of selective reflection spectrum (blue dots) recorded for $B\approx 500\;$G together with the equivalent theoretical spectrum calculated for $\ell=750\;$nm (orange solid line), see the text.}
    \label{fig:Fig4}
\end{figure}

To validate the theoretical model, we have recorded and analyzed experimental SR spectra from a NC obtained for magnetic field in a range of 450 to 800\;G. The results are summarized in Fig.\;\ref{fig:Fig4}, were we plot the extracted positions of the $F= 3 \rightarrow 5'$ MI transitions as a function of the applied magnetic field. The recorded positions of these transitions are seen to agree with the predictions following the model described in Sec.\;\ref{sec:theory}. Note that for $B>650\;$G, the mode-hop free frequency scanning range of the laser greatly deteriorated at the time the data was acquired for these fields. For this reason, we were not able to record all the seven $F=3\rightarrow5'$ MI transitions.

The top inset in Fig.\;\ref{fig:Fig4} shows an experimental SR spectrum (blue dots) processed with first order derivative (dSR) and obtained for $B \approx 500\;$G. The orange solid line shows the equivalent theoretical spectrum obtained with $\ell = 750\;$nm and $B=500\;$G. Details on how to calculate dSR spectrum can be found e.g. in Refs.\;\cite{dutierJOSAB2003,sargsyanJPhysB2025}. Note that, for fields below 2.5\;kG, $F=3\rightarrow5'$ MI transitions partially overlap with the group of transitions $F=3\rightarrow4'$ for which one transition can be observed between \textcircled{7} and \textcircled{6} (another one completely overlaps with \textcircled{6} for this magnetic field).

\section*{Conclusion}

In conclusion, we have studied, both experimentally and theoretically, the magnetic dependence of the high frequency wing spectrum of the $6^2\text{S}_{1/2}\rightarrow7^2\text{P}_{3/2}$ transition observed under right-hand circularly-polarized laser light at 456\;nm. Our studies have focused on seven $F=3\rightarrow5'$ magnetically-induced transitions. Among the important features of these MI transitions is their large frequency shift with respect to the unperturbed hyperfine transitions, which reaches $\sim 17~\mathrm{GHz}$ in magnetic fields of $\sim 3~\mathrm{kG}$, easily achievable with permanent magnets \cite{PoncianoOjeda2020}. Besides, they reach their maximum intensity in the magnetic field range of $0.2$--$3~\mathrm{kG}$ where they are more intense than many conventional ($\Delta F = 0,\pm1$) atomic transitions. 

These results are important for both fundamental studies and practical applications, particularly for the development of magnetometers with sub-micron spatial resolution in the blue spectral region.
For example, these seven MI transitions may be used 
to form EIT in a three-level Ladder system and single out different excitation paths to the $32$S Rydberg state of Cs atoms with 456 and 1070\;nm lasers \cite{Urvoy2016_PhD}.
Besides, as demonstrated in Ref.\;\cite{tonoyan2023formation}, using a nanocell with a thickness $\ell = 100~\mathrm{nm}$ and 
$F=1\rightarrow3'$ MI transitions of $^{87}$Rb D$_2$ line,
it is possible to realize a magnetometer with a spatial resolution of 
$1\;\mu\mathrm{m}$. This is particularly important for mapping magnetic fields with strong spatial 
gradients exceeding $3\,\mathrm{G}/\mu\mathrm{m}$~\cite{liang2020}. Finally, these transitions may be used to create an optical frequency reference operating at controllable and highly shifted frequencies (with respect to that of the unperturbed hyperfine transition) in the $\sim 450~\mathrm{nm}$ blue spectral region.


\section*{Author Contributions}

Armen Sargsyan: Writing -- review \& editing, Writing -- original draft, Investigation, Funding acquisition, Data curation.

Arevik Amiryan: Writing -- review \& editing, Writing -- original draft, Visualization, Software, Formal analysis.

Emmanuel Klinger: Writing -- review \& editing, Writing -- original draft, Visualization, Software, Formal analysis.

David Sarkisyan: Writing -- review \& editing, Writing -- original draft, Supervision, Investigation, Funding acquisition.

\section*{Funding}
 AS and DS acknowledge financial support from the Higher Education and Science Committee of the MESCS RA under the project No 25RG-1C008.

\section*{Declaration of Competing Interest}




All authors declare that they have no known competing financial interests or personal relationships that could have appeared to influence the work reported in this paper.


\section*{Disclosures}

LLM artificial intelligence agents were used to assist the editing of this manuscript, but all scientific results, concepts, and claims
represent original work by the human authors listed.

\section*{Data availability statement}
The data that support the findings of this study are available from the corresponding author upon reasonable request.

\section*{References}
\bibliography{bib-new}

\end{document}